\documentclass[lettersize,journal]{IEEEtran}
\usepackage{amsmath,amsfonts}
\usepackage{algorithmic}
\usepackage{algorithm}
\usepackage{array}
\usepackage[caption=false,font=normalsize,labelfont=sf,textfont=sf]{subfig}
\usepackage{textcomp}
\usepackage{stfloats}
\usepackage{url}
\usepackage{verbatim}
\usepackage{graphicx}
\usepackage{cite}

\begin{document}

\title{Towards Robust and Interpretable EMG-based Hand Gesture Recognition using Deep Metric Meta Learning}

\author{Simon Tam,
Shriram Tallam Puranam Raghu,
Étienne Buteau,
Erik Scheme, \IEEEmembership{Senior Member, IEEE},
Mounir Boukadoum, \IEEEmembership{Life Senior, IEEE},
Alexandre Campeau-Lecours, \IEEEmembership{Member, IEEE},
Benoit Gosselin, \IEEEmembership{Member, IEEE}
        % <-this % stops a space
\thanks{This work was supported in part by the Natural Sciences and
Engineering Research Council of Canada (NSERC) Alliance Grant
ALLRP 570710-2021 and in part by the Canada Research Chair in
Smart Biomedical Microsystems, the Microsystems Strategic Alliance
of Quebec (ReSMiQ), and the Quebec Research Funds in Science and
Technologies (FRQNT) grant number B2X-335236.}% <-this % stops a space
}



\maketitle

\begin{abstract}
  Current electromyography (EMG) pattern recognition (PR) models have been shown to generalize poorly in unconstrained environments, setting back their adoption in applications such as hand gesture control. This problem is often due to limited training data, exacerbated by the use of supervised classification frameworks that are known to be suboptimal in such settings. In this work, we propose a shift to deep metric-based meta-learning in EMG PR to supervise the creation of meaningful and interpretable representations. We use a Siamese Deep Convolutional Neural Network (SDCNN) and contrastive triplet loss to learn an EMG feature embedding space that captures the distribution of the different classes. A nearest-centroid approach is subsequently employed for inference, relying on how closely a test sample aligns with the established data distributions. We derive a robust class proximity-based confidence estimator that leads to a better rejection of incorrect decisions, i.e. false positives, especially when operating beyond the training data domain. We show our approach's efficacy by testing the trained SDCNN's predictions and confidence estimations on unseen data, both in and out of the training domain. The evaluation metrics include the accuracy-rejection curve and the Kullback-Leibler divergence between the confidence distributions of accurate and inaccurate predictions. Outperforming comparable models on both metrics, our results demonstrate that the proposed meta-learning approach improves the classifier's precision in active decisions (after rejection), thus leading to better generalization and applicability.
\end{abstract}


\section{Introduction}
\IEEEPARstart{H}{}and gesture recognition (HGR) with electromyography (EMG) involves recognizing user intent by analyzing the complex muscle activation patterns generated when different gestures are elicited. Researchers have employed machine and deep learning (ML, DL) classification models to learn and identify the gestures from these intricate signals, which are subsequently used to control devices \cite{gopal2022, parajuli_real-time_2019}. DL, in particular, has attracted a lot of attention thanks to its ability to learn relevant features directly from the data, avoiding the need for feature engineering \cite{jiang2023}.

While researchers have reported high recognition accuracy in controlled environments, the performance of the models degrades when deployed in unconstrained real-word environments \cite{campbell2020}. This is generally caused by external noise sources, interference, and confounding factors such as limb orientation and sensor displacement relative to the limb. These factors ultimately result in poor user experience, thus motivating researchers to seek methods to mitigate the performance degradation arising from them. While acquiring larger EMG datasets to encompass a broader test domain may lead to better generalization, it is not practical due to the time and effort required from the end users to provide such data. Within the constraint of limited training data, an effective method to improve usability is the use of decision rejection using confidence metrics (e.g., posterior probabilities) obtained from the model \cite{robertson2019, scheme2015, shri2023}. However, this scheme assumes that the resulting confidence values can faithfully be interpreted as measures of class membership. This presents a significant challenge, especially in the case of DL models, as they typically function as black boxes, complicating the interpretation of machine decisions, let alone confidence levels \cite{tjoa2021}. Moreover, it is well known that DL models trained with the conventional cross-entropy loss are generally poorly calibrated \cite{pmlr-v70-guo17a}, tending to be overconfident even when the decisions are incorrect. This poses a notable challenge in safety-critical systems like EMG-controlled prostheses, where the repercussions of erroneous decisions far outweigh those of inaction. Such errors could potentially result in hazardous movements of the prosthesis, posing risks to both the user and bystanders. Researchers have argued that overconfidence is a fundamental problem with supervised classification frameworks, and have explored ways to better calibrate the networks \cite{NEURIPS2019_36ad8b5f}. However, this still does not help solve the model interpretability issues. 

In this work, we aim to address the intrinsic limitations of the conventional classification frameworks in EMG pattern recognition (PR) when it comes to model generalization, interpretation, and usability. We do so by framing EMG PR as a \emph{representation learning problem} rather than a conventional classification problem. To this effect, we present a deep metric-based meta-learning framework. At the front end of the model is a Siamese Deep Convolutional Neural Network (SDCNN) that is trained with a contrastive triplet loss to learn a semantically meaningful Euclidean embedding space. At the back end, a nearest centroid (NC) classifier is then employed to use contextual distance-based information in the learned embedding space for inference and confidence estimation. This approach leverages the feature extraction capabilities of deep learning while incorporating an element of interpretable AI through the transparent and intuitive nature of the NC classifier. We evaluate the model under 3 EMG PR test scenarios: 1) in-domain predictions, 2) domain-divergent predictions, 3) out-of-domain predictions. Among other evaluation metrics, we use the accuracy-rejection curve (ARC) and the Kullback-Leibler (KL) divergence between confidence distributions of accurate and inaccurate predictions. The ARC provides insights into the trade-off between accuracy and the rejection of uncertain predictions, while the KL divergence measures the difference in confidence distributions between correct and incorrect predictions, shedding light on the model's calibration and decision-making process. Thanks to meta-learning and NC classification, the proposed approach outperforms comparable methods on those metrics, offering a deep learning approach that is better suited to deal with the generalization problem caused by inherent data limitations in EMG PR. With improved confidence estimation to inform decision rejection, it also contributes to build more efficient DL models for EMG-based human-machine interfaces (HMI).

\begin{figure*}[t]
\begin{center}
\centerline{\includegraphics[width=\textwidth]{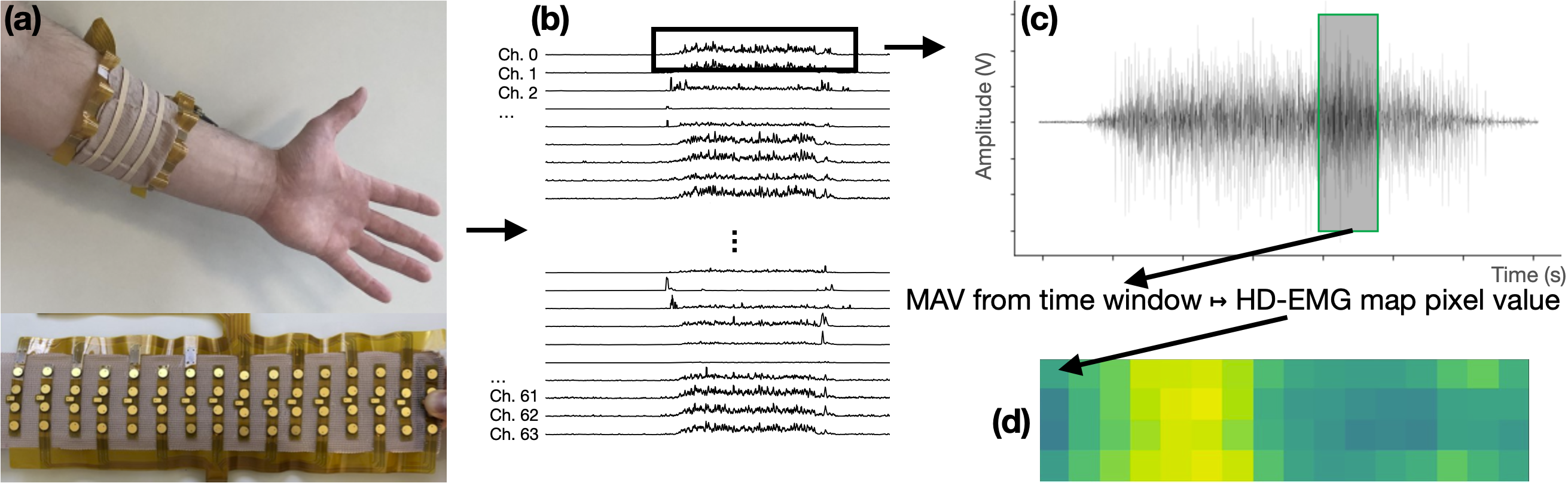}}
\caption{High-density electromyography (HD-EMG) muscle activity mapping. (a) Non-invasive HD-EMG array of electrodes. (b) Time-series signal are individually processed if needed. This example shows 64 channels for corresponding $4 \times 16 = 64$ electrodes. (c) From each channel, a mean absolute value (MAV) window captures a measure of the muscle contraction intensity. (d) The MAV of each channel is mapped to a pixel value in the muscle activity heat map, and the pixel location corresponds to the electrode's physical position in the array.}
\label{fig:hdemg_mapping}
\end{center}
\end{figure*}

\section{Related Work}
This section presents the related work in 3 areas: EMG pattern recognition, decision rejection, and representation learning.
\paragraph{Electromyography pattern recognition}
Classification-based pattern recognition is the leading method for EMG HGR \cite{jaramillo2020}. Because of the user-specific nature of EMG, EMG-driven HMIs require end user training data. However, obtaining enough data to encompass all possible variations in movements is impractical as it puts extraordinary burden on the users. Researchers thus only collect a few repetitions of data from selected gestures, typically when contractions are maintained without any dynamic movements (also called static or stationary data). This constraint results in models performing sub-optimally due to limited training. To alleviate this, recent works have proposed approaches such as transfer learning \cite{tam2021, cote2019}, domain adaptation \cite{cote2021, du2017}, and data augmentation schemes \cite{tsinganos2020, chamberland2023}. While these methods reinforce training for a given set of gesture classes, the core model still faces generalization issues outside of the trained scope as users inevitably generate loosely-controlled EMG patterns in unconstrained test scenarios.

\paragraph{Decision rejection}
In cases when device inaction is preferred to prediction error, decision rejection can reduce the percentage of false positive activations, albeit affecting some true positives in the process. In the DL space, \cite{bao2022} proposed the use of CNN posterior probabilities to model a confidence estimator for decision rejection. In \cite{lin2022}, it is proposed to replace the softmax probabilities in a CNN with evidence-based output activation and loss functions to embed uncertainty estimates in the output. However, these approaches rely on the quality and variability of the training dataset for the estimators, thus limiting their usefulness. Interpretability still remains limited, as the \emph{black box}-type learning of the downstream training loss does not yield intelligible features for users. Wu \textit{et al.} \cite{wu2021} use a metric center loss to train the last hidden layer of their model in conjunction with the conventional cross-entropy loss for a CNN model. Additionally, an autoencoder (AE) is trained for each class and novel samples are rejected based on the reconstruction error of the AEs. Though this approach steps in the right direction to address the generalization issues of cross-entropy loss, it has two weaknesses. First, along with the base model, one AE is required for \emph{every} gesture class that is being considered; second, the performance of the AEs may still rely on the training data for proper generalization, which limits the benefits of the approach when data availability is constricted.

\paragraph{Representation and distance learning}
Meta-learning has been used to solve challenges of data availability and model generalization. In deep learning in particular, it can add \textit{good} inductive bias in the learning process \cite{huisman2021}. Specific implementations may use metric, model, or optimization-based methods. The exploration of meta-learning in EMG applications is, however, under-explored in comparison to other fields. Prorokovi{\'c} \textit{et al.} \cite{prorokovi2020} use model-agnostic meta-learning (MAML) to facilitate model adaptation to EMG signal variations over time. In \cite{rahimian2021}, a few-shot learning (FSL) framework is used to maximize learning and generalization with limited data. Fan \textit{et al.}\cite{fan2022} use a similar approach, but also implement MAML to specifically facilitate adaptation of EMG pattern recognition systems to new users. The previous works focused on achieving better recognition accuracy under different conditions, but largely ignored interpretability. In this work, we use metric-based meta-learning from a different angle, with supervised representation learning as a way to generate interpretable and actionable signals from the model, such as confidence estimation to inform automatic decision rejection.

\begin{figure*}[t]
\begin{center}
\centerline{\includegraphics[width=0.9\textwidth]{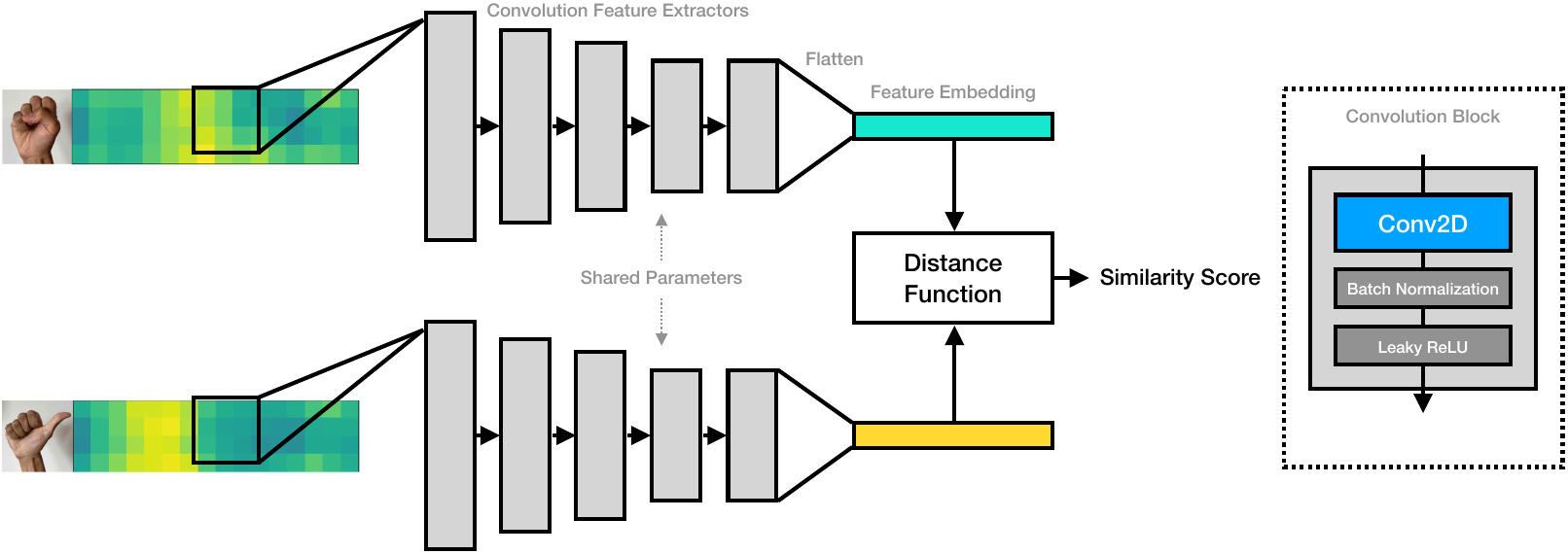}}
\caption{Siamese deep convolutional neural network architecture. The two branches of 2D convolution feature extractors embed the input data through multiple convolution layer blocks and yield a flattened feature vector. The feature vectors are then compared with a distance function to assess similarity. More than 2 branches may be used to compare larger tuples of input data.}
\label{fig:sdcnn}
\end{center}
\end{figure*}

\section{Methods}
\subsection{High-density electromyography signals and data}

In this work, we employ a high-density electromyography (HD-EMG) array consisting of 64 electrodes organized as a $4 \times 16$ grid \cite{chamberland2023, tam2019}. As depicted in Fig. \ref{fig:hdemg_mapping}, the HD-EMG can be interpreted as heat-map images representing the intensity of muscle contraction in the limb. With periodic sampling, sequences of HD-EMG data are produced with a shape of (N, H, W), where \textit{N} is the time/sample axis, and \textit{H} and \textit{W} correspond to the height and width of the sensor array. This HD-EMG image representation enables direct use of 2D convolutions to extract spatial correlation patterns.

\subsection{Siamese deep convolutional neural network for representation learning}

To extract informative features from the HD-EMG data, we employ a Siamese Deep Convolutional Neural Network (SDCNN) architecture. Siamese networks are sometimes also referred to as twin, matching, or prototypical networks \cite{huisman2021}. As shown in Fig. \ref{fig:sdcnn}, the parallel branches of the SDCNN consist of 2D convolution blocks, with shared parameters between the branches. Details of the architecture are presented in Table \ref{tab:nnlayers}. Akin to AlexNet \cite{krizhevsky2017}, the convolution kernels are progressively smaller along the depth of the network. Given the relatively low image resolution of HD-EMG maps (compared to conventional images), downsampling the feature maps with pooling layers was deemed unnecessary. After the convolution blocks, the last feature maps are flattened to form the feature embedding vector.

We train the network with a triplet loss function \cite{schroff2015}, which guides the network to learn a feature embedding space that enforces proximity between sampled same-class vectors, and maximizes inter-class distance. New samples can then be projected into the embedding space, which will enable greater understanding of the sample's membership to different class clusters. The loss is mathematically defined by:
\begin{center}
    $loss_{triplet} = max(d_{same} - d_{diff} + \alpha, 0)$,
\end{center}
where, for a given data point, $d_{same}$ is its Euclidean distance to a same class example, $d_{diff}$ is its distance to a different class example, and $\alpha$ is a margin parameter. In this work, we use the semi-hard online implementation \cite{schroff2015}, where the sampled triplets from the training set batches are subject to the following constraint:
\begin{center}
    $||f(x_i^{ref}) - f(x_i^{same})||_2^2 < ||f(x_i^{ref}) - f(x_i^{diff})||_2^2 < ||f(x_i^{ref}) - f(x_i^{same})||_2^2 + \alpha$,
\end{center}
where $x_i^{ref}$ is the reference (anchor) sample, $x_i^{same}$ is another sample from the same class, $x_i^{diff}$ is another sample from a different class, and $\alpha$ is the configurable margin parameter. This makes the training more efficient by avoiding trivial triplets that would yield negative losses. The $\alpha$ margin also enforces the sub-selection of semi-hard triplets ($x_i^{diff}$ close to $x_i^{ref}$, but still farther than $x_i^{same}$). This process also excludes so-called hard triplets which may impair training convergence.

\begin{table}[t]
  \caption{Layers of the convolution feature extractors in the twin channels of the SDCNN}
  \label{tab:nnlayers}
  \centering
  \begin{tabular}{|l|l|l|l|l|} \hline
    Layer & Hyper-param. & Add. layers & Activation \\ \hline \hline
    Conv2D    & 32, 13x13, 0-padding & Batch norm. & Leaky ReLU (0.01)\\ \hline
    Conv2D    & 32, 9x9, 0-padding & Batch norm. & Leaky ReLU (0.01)\\ \hline
    Conv2D    & 32, 5x5, 0-padding & Batch norm. & Leaky ReLU (0.01)\\ \hline
    Conv2D    & 32, 3x3, 0-padding & Batch norm. & Leaky ReLU (0.01)\\ \hline
    Conv2D    & 8, 3x3, no padding & Batch norm. & Leaky ReLU (0.01)\\ \hline
    Flatten      & N/A & N/A & N/A\\ \hline
  \end{tabular}
\end{table}

\subsection{Nearest centroid classifier and decision confidence estimation}

Once the SDCNN learns to embed HD-EMG data into a semantic Euclidean feature space, multi-class discrimination can be done with a nearest centroid approach. The centroid prototypes for each class are obtained by computing the mean feature embedding vectors from the training data. An added benefit of Siamese networks and the NC approach is the utilization of few-shot learning, which enables training with minimal quantities of examples, and subsequent handling of new prototypes without having to retrain the neural network \cite{wang2022}.

For classification, the incoming test data is attributed to the nearest centroid's class. Because the SDCNN learns the feature space based on EMG data similarity, distance from the other classes can be used to provide contextual information along with the prediction and inform a confidence estimator. The distance to each prototype is turned into a class membership score with the softmax function as in:
\begin{equation}
    C = softmax([D_0, D_1, ..., D_{N-1}])
\end{equation}
\begin{equation}
    D_c = 1 - \frac{d_c}{\sum_{i=0}^N d_i}
\end{equation}
where $d_c$ represents the distance to the class $c$ centroid, and $N$ the total number of classes. The softmax scores relay how confident the decision (highest score) is relative to the other classes. The same cannot be said of classification neural networks using a cross-entropy loss, where one-hot encoded target outputs don't model class similarities and lead to overconfident predictions \cite{nguyen2015}.

%\section{Experimental setup}
\subsection{Dataset}
We compiled an HD-EMG dataset collected from 10 able-bodied users in accordance with relevant guidelines, regulations, and experimental protocols, as approved by the Laval University Research Ethics Committee (approbation number: 2019-268 A-1 R-3 / 23-11-2022). Informed consent was obtained from all participants. We used the 64-channel wearable sensor (4x16 array around the forearm) presented in \cite{chamberland2023}, using a 1 kHz sampling frequency and 16-bit analog-to-digital resolution to record the data from each participant.

\begin{figure}[t]
\begin{center}
\includegraphics[width=0.9\columnwidth]{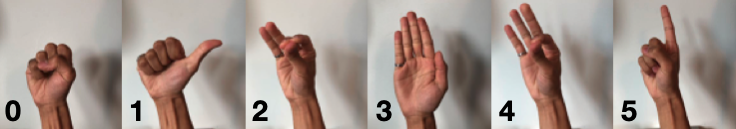}
\caption{Gesture classes used for classification: 0) close fist, 1) thumbs up, 2) chuck grip, 3) rest, 4) fine pinch, 5) index extension.}
\label{fig:gestures}
\end{center}
\end{figure}

\paragraph{Static hand gesture data}
A static set was created by having each participant perform 6 hand gestures, pictured in Fig. \ref{fig:gestures}. Each gesture was held steady for 5 seconds, and repeated 10 times. Only the steady-state portion of the hand gestures were recorded.

\paragraph{Dynamic hand gesture data}
For each participant, a dynamic sequence was also recorded to include transitions between gestures in the data. In a continuous recording, the participant cycled through the 5 active hand gestures, holding them for 5 seconds each, and resting in neutral hand position for 5 seconds in-between. The users were visually cued with progress bars to indicate when to switch gesture states.

\paragraph{Signal processing}
The analog EMG signals were band-pass filtered with a low cutoff frequency of 20 Hz (1st-order characteristic) and a high cutoff of 300 Hz (3rd-order Butterworth characteristic). 60 Hz power line interference was removed from the digitized signals using a second-order infinite impulse response (IIR) notch filter with a quality factor, Q, of 30. For each channel's time-series signals, \textit{DC offset removal} was implemented as a 100 ms moving average subtraction, and \textit{envelope smoothing} was done with a 100 ms mean absolute value (MAV) filter to enhance subsequent samples correlation in the 2D muscle-map representation \cite{tam2019}. For the static dataset, the processing time windows for the MAV and offset removal were non-overlapping (window increment of 100 ms) to reduce training computation. A pilot experiment showed no appreciable difference between using 1 ms and 100 ms increment because the overlap created by shorter strides only added redundancy between the data points, providing no additional information during training. For the dynamic sequences, however, the processing time windows were made to overlap, with increments of 1 ms, to better capture the dynamics during gesture transitions.

\subsection{Reference models}
\paragraph{Deep Convolutional Neural Network}
To verify the efficacy of the proposed metric learning framework, a baseline DCNN model was implemented using the exact same convolution layers as the SDCNN (Table \ref{tab:nnlayers}). A classification head was appended to the model using a 128-unit fully-connected (FC) layer, with a leaky ReLU activation (negative slope of 0.01), going into the \emph{C}-unit FC output layer, where \emph{C} is the number of classes in the training dataset. During training, 50\% dropout was used after the 128-unit FC layer. This model was trained in a typical classification framework using the categorical cross-entropy loss. Given similar model complexity in terms of parameters, the contribution of the different learning frameworks can be directly compared. The confidence estimation of the DCNN model was extracted using the predictive probabilities of the softmax output activation. 

\paragraph{Support Vector Machine}
To also provide a non-DL baseline, we implemented a support vector machine (SVM) classifier, which has been shown to outperform other common classifiers in confidence-based decision rejection \cite{scheme2015}. The employed SVM model uses a radial basis function (RBF) kernel with parameters \emph{C} of 550 and \emph{gamma} of 5e-08, as tuned for best accuracy in a pilot study with 1 user and 5-fold cross-validation. The typical non-DL methods require feature engineering prior to classification; in our case, we present the same preprocessed data to the SVM as with the other models, but in a flattened vector of 64 elements (as opposed to the 4 by 16 array input representation for the 2D convolution models). Therefore, it effectively sees the MAV feature for each of the 64 channels for classification purposes. A variety of other features have been studied in EMG PR \cite{phinyomark2013}, but a uniform preprocessing stage was preferred in this work for comparison purposes. During training, we obtained the posterior probabilities and used them as confidence scores for the SVM model.

\paragraph{CNNSC}
As a way to extend the discriminative property of the features learned by the CNN, a center loss function can be added to the traditional softmax loss \cite{wu2021}. This additional loss is measured using the features generated by the penultimate dense layer of the DCNN model. This loss function aims to penalize discrepancies between learned features and their respective centers, encouraging samples belonging to the same gesture to converge towards their central point in the embedding space. The joint loss uses an hyperparameter $\tau$ to help balance the two losses. The authors claim that the softmax loss maximizes the distance between different classes in the embedding space, while the center loss minimizes the distance within the same class. Also for this approach, the model's confidence estimation is derived from the softmax output activation.

\paragraph{ECNN}
To evaluate the ability of the different models to represent a measure of uncertainty compared to the SDCNN, we implemented an Evidential Convolutional Neural Network (ECNN) \cite{lin2022} that extends the baseline DCNN model. Instead of using softmax to directly extract class probabilities, ECNN substitutes it with an activation layer like ReLU or SoftPlus to generate evidence vectors that are passed through a Dirichlet probability density function. The network is trained using a custom sum-of-squares loss function that integrates an additional Kullback-Leibler (KL) divergence term with a trade-off coefficient $\lambda$ that helps regulate the Dirichlet distribution. With the proposed ECNN framework, four uncertainty measures are introduced: vacuity, dissonance, entropy and negative maximum probability. The vacuity and dissonance are specific to ECNN and represent its evidential uncertainty, describing uncertainty due to lacking evidence and uncertainty due to conflicting evidence. Regarding the classification accuracy, using the negative maximum probability as an uncertainty metric provided the best scores. It was thus used to obtain the confidence scores for decision thresholding. The model was trained using a ReLU function as the substitutes activation layer and a $\lambda$ coefficient of 0.1.

\subsection{Experiments}

% \subsection{Training details}
For all experiments, the model training, validation, and testing were completed for each user separately. Unless specified otherwise, a per-trial leave-one-out cross-validation (LOOCV) scheme was used for training and evaluation on the static hand gesture dataset. With each gesture class repeated 10 times (i.e. 10 trials), the first trial was set aside for validation, and the remaining 9 were used for the LOOCV. The four neural network models used a training batch size of 128, the Adam optimizer with a learning rate of $10^{-3}$ (SDCNN, DCNN) or $10^{-4}$ (CNNSC, ECNN), and early stopping with best epoch recall to halt training after 5 epochs of validation loss improvements under $10^{-4}$. The triplet loss $\alpha$ parameter for the SDCNN is set to 20.0. The $\tau$ parameter for the CNNSC was set to $5*10^{-5}$. For dataset consistency across the model comparison, the SVM used the same training/test CV folds, thus disregarding the validation trial. The main frameworks and versions used are: TensorFlow v2.12.0, TensorFlow Addons v0.20.0, scikit-learn v1.2.2.

To evaluate the usefulness of the distance metric-based confidence estimation, 3 offline experiments were conducted to assess model performance in different test data conditions: in-domain, domain-divergent, and out-of-domain. Then, to showcase the expected real-time performance, the models' behaviors were compared during a dynamic gesture sequence featuring in-domain and domain-divergent segments.

\subsubsection{In-domain misclassification rejection}
This test assessed the ability to discriminate between accurate and inaccurate in-domain model predictions based on each model's internal confidence estimation. The test data were sampled from the same domain as the training data, i.e. EMG signals from static contractions of the same gesture classes.

\subsubsection{Domain-divergent transient-induced error rejection}
The domain-divergent test assessed the ability to identify predictions on data falling just outside the training data domain, due to confounding factors at the source. In this case, the test data consisted of the dynamic EMG sequences. With the training domain consisting of static contractions, the test domain diverged due to the inclusion of EMG data from gesture transitions \cite{shri2022}. These transitions cause uncertainty and have been shown to induce prediction errors due to domain divergence. We thus evaluate the models' abilities to discriminate in-domain from domain-divergent predictions for decision rejection purposes.

\subsubsection{Out-of-domain unforeseen class rejection}
The out-of-domain test assesses the ability to identify predictions made on data unrelated to the training domain. For this experiment, we introduce gesture classes in the test data that were not part of the training classes ensemble. Predictions on unknown/unforeseen classes are inherently inaccurate and should thus lead to lower confidence than in-domain predictions.

For this experiment alone, the LOOCV was performed class-wise rather than trial-wise, setting aside each of the 6 classes from the training data. For each fold, the first contraction trial of each training class was used for validation, the last two for in-domain reference testing, and the rest for training. The left-out data (10 trials of the left-out class) were then passed as test data to obtain the out-of-domain predictions and confidence scores. Each fold thus consisted of a ratio of 50\% in-domain/out-of-domain trials (2 trials $\times$ 5 known classes and 10 trials of the left-out class).

\begin{figure}[t]
\begin{center}
    \includegraphics[width=0.987\columnwidth]{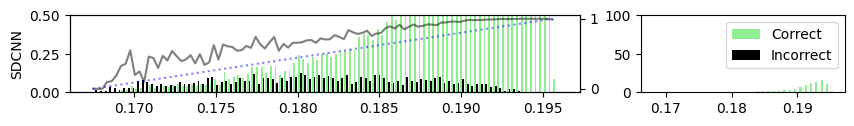}
    \vspace{-1mm}
    \includegraphics[width=1\columnwidth]{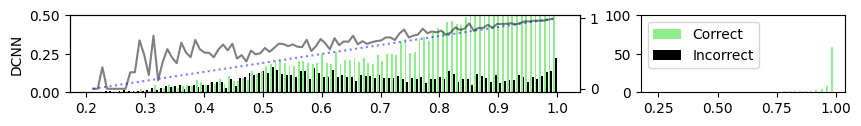}
    \vspace{-1mm}
    \includegraphics[width=1\columnwidth]{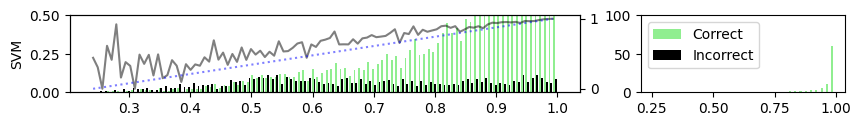}
    \vspace{-1mm}
    \includegraphics[width=0.997\columnwidth]{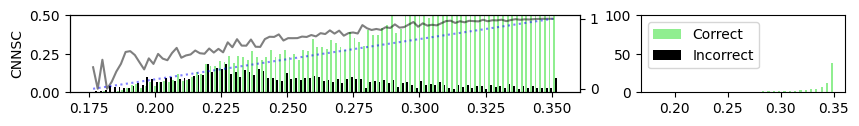}
    \vspace{-1mm}
    \includegraphics[width=1\columnwidth]{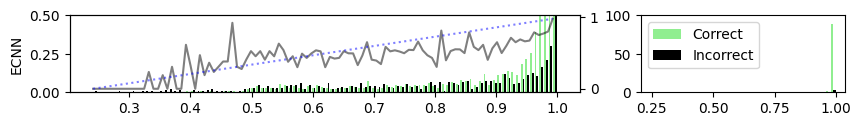}
    \vspace{-5mm}
\caption{In-domain test confidence score distribution. The vertical axes represent the percentage of total test predictions. Close-up view with 100 bins, overview with 35 bins. Calibration curves, representing the percentage of correct predictions for each bin, are overlaid (secondary Y-axis), with linear trend shown as reference for ideal calibration (dotted line).}
\label{fig:id_distributions}
\end{center}
\end{figure}

\section{Results}
\begin{figure}[t]
\begin{center}
    \includegraphics[width=0.987\columnwidth]{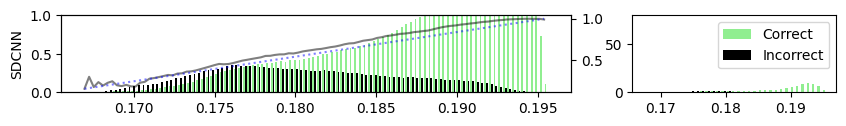}
    \vspace{-1mm}
    \includegraphics[width=1\columnwidth]{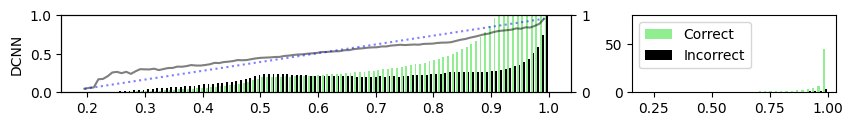}
    \vspace{-1mm}
    \includegraphics[width=1\columnwidth]{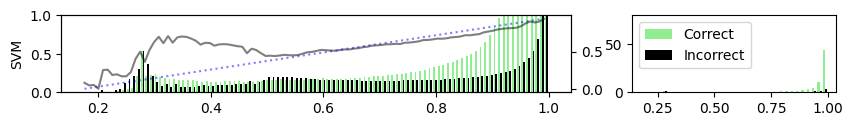}
    \vspace{-1mm}
    \includegraphics[width=0.997\columnwidth]{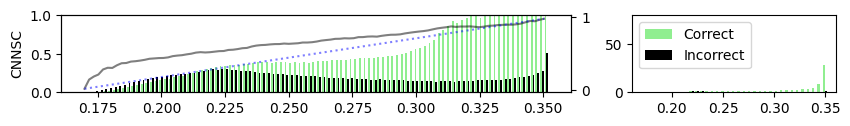}
    \vspace{-1mm}
    \includegraphics[width=1\columnwidth]{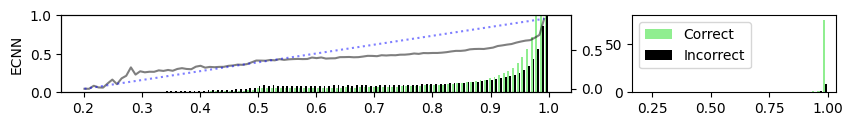}
    \vspace{-5mm}
\caption{Domain-divergent test confidence score distribution. The vertical axes represent the percentage of total test predictions. Close-up view with 100 bins, overview with 35 bins. Calibration curves are overlaid (secondary Y-axis), with linear trend shown as reference.}
\label{fig:dd_distributions}
\end{center}
\end{figure}

\begin{figure}[t]
\begin{center}
    \includegraphics[width=0.987\columnwidth]{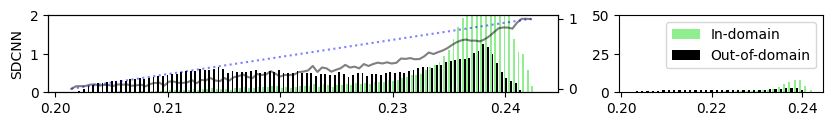}
    \vspace{-1mm}
    \includegraphics[width=1\columnwidth]{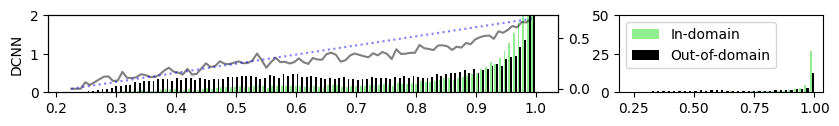}
    \vspace{-1mm}
    \includegraphics[width=1\columnwidth]{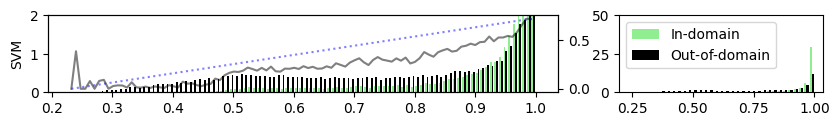}
    \vspace{-1mm}
    \includegraphics[width=0.997\columnwidth]{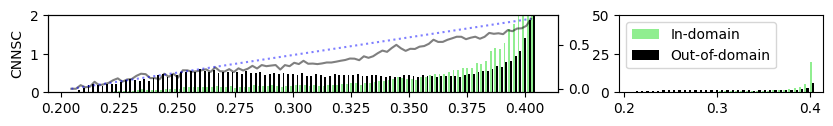}
    \vspace{-1mm}
    \includegraphics[width=1\columnwidth]{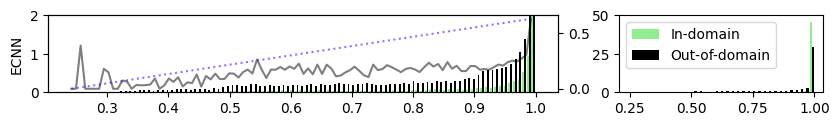}
    \vspace{-5mm}
\caption{In \& out-of-domain test confidence score distribution. The vertical axes represent the percentage of total test predictions. Close-up view with 100 bins, overview with 35 bins. Calibration curves are overlaid (secondary Y-axis), with linear trend shown as reference.}
\label{fig:ood_distributions}
\end{center}
\end{figure}

\begin{figure*}
    \centering
  \subfloat[ROC, in-domain test.\label{fig:id_roc}]{%
       \includegraphics[width=0.3\linewidth]{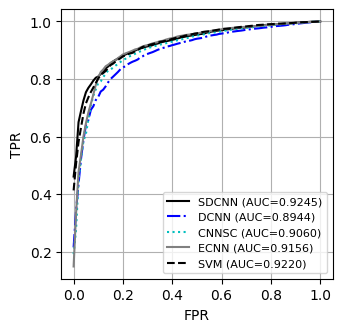}}
    \hfill
  \subfloat[ROC, domain-divergent test.\label{fig:dd_roc}]{%
        \includegraphics[width=0.3\linewidth]{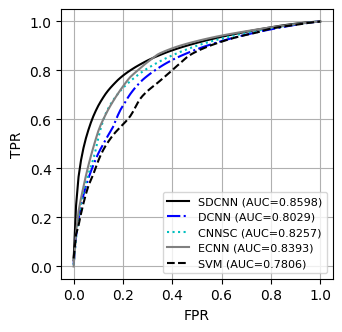}}
    \hfill
  \subfloat[ROC, out-of-domain test.\label{fig:ood_roc}]{%
        \includegraphics[width=0.3\linewidth]{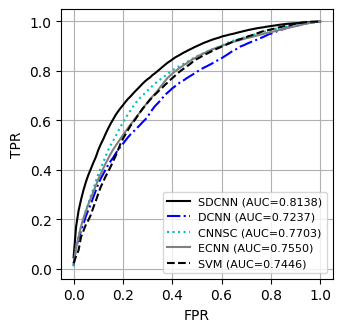}}
    \\
  \subfloat[PRC, in-domain test.\label{fig:id_prc}]{%
       \includegraphics[width=0.3\linewidth]{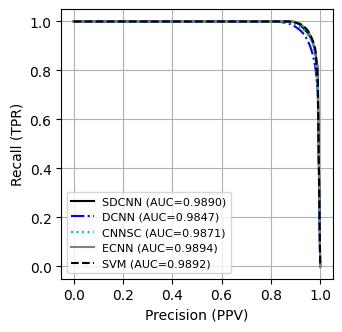}}
    \hfill
  \subfloat[PRC, domain-divergent test.\label{fig:dd_prc}]{%
        \includegraphics[width=0.3\linewidth]{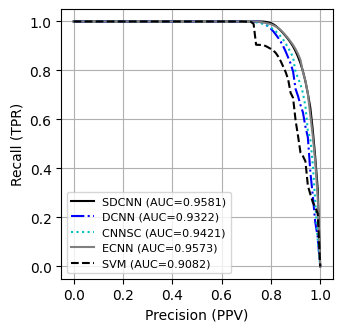}}
    \hfill
  \subfloat[PRC, out-of-domain test.\label{fig:ood_prc}]{%
        \includegraphics[width=0.3\linewidth]{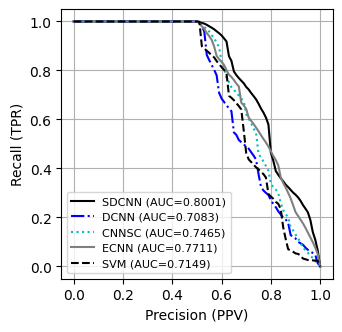}}
    \\
  \subfloat[ARC, in-domain test.\label{fig:id_arc}]{%
       \includegraphics[width=0.3\linewidth]{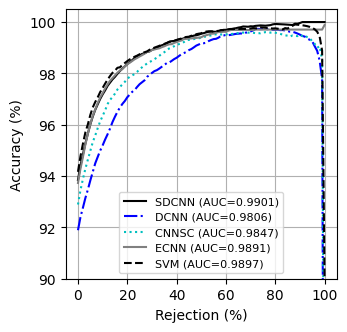}}
    \hfill
  \subfloat[ARC, domain-divergent test.\label{fig:dd_arc}]{%
        \includegraphics[width=0.3\linewidth]{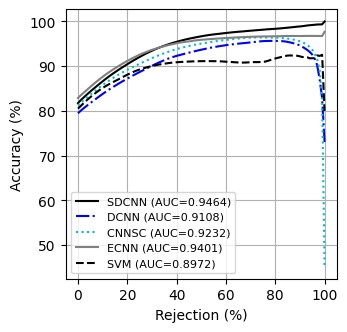}}
    \hfill
  \subfloat[ARC, out-of-domain test.\label{fig:ood_arc}]{%
        \includegraphics[width=0.3\linewidth]{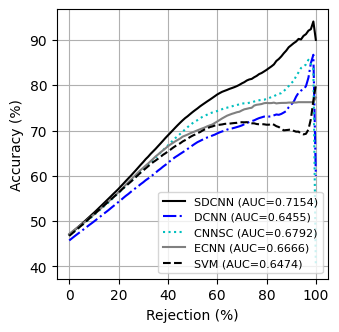}}
    \\
  \caption{Receiver operating characteristic, precision-recall and accuracy-rejection curves (ROC, PRC, ARC) for confidence-based decision rejection. The ARC shows the trade-off between the amount of rejection and the active accuracy (classification accuracy among the non-rejected predictions).}
  \label{fig:curves} 
\end{figure*}

Where applicable, each experiment's test accuracy is presented in Table \ref{tab:results}. For each experiment case, the distributions of confidence scores for decision rejection can be observed on Fig. \ref{fig:id_distributions}, \ref{fig:dd_distributions}, and \ref{fig:ood_distributions}. The KL divergences ($D_{KL}$), or relative entropy, are reported in Table \ref{tab:results} as a measure of probability distribution difference. To evaluate the confidence-based decision rejection ability, the curves for the receiver operating characteristic (ROC), precision-recall (PRC), and accuracy-rejection (ARC) are displayed in Fig. \ref{fig:curves}, with their area under the curve (AUC) reported in Table \ref{tab:results}. The curves and AUC are calculated for each user independently, and then averaged.

\subsection{In-domain misclassification rejection}
The in-domain test evaluated performance on the LOOCV test folds. The test accuracy is presented in Table \ref{tab:results}. After the LOOCV, all test predictions and their confidence scores are aggregated and relabeled as true or false based on whether they were accurate or not, with their distributions shown in Fig. \ref{fig:id_distributions}. The $D_{KL}$, in this case, quantifies how the \emph{incorrect} distribution differs from the \emph{correct} one.

All models performed well in this test condition, with accuracies $>90\%$ corroborating the existing body of work \cite{li2021}. The SVM had the highest accuracy, SDCNN had the best $D_{KL}$, AUROC, and AUARC; while the ECNN had the best AUPRC. However, none of the metrics were significantly different across the different classifiers in this case.

\subsection{Domain-divergent transient-induced error rejection}
In this experiment, the LOOCV test folds were used as a reference for baseline model accuracy, and referred to as \emph{reference accuracy} in \ref{tab:results}. For each fold, the domain-divergent test accuracy was calculated on the dynamic EMG sequence, for each user respectively. For decision rejection, all dynamic test predictions were relabeled as true or false based on whether they were accurate or not, with their confidence score distributions displayed in Fig. \ref{fig:dd_distributions}. The $D_{KL}$ measures how much the \emph{incorrect} distribution differs from the \emph{correct} one.

The overall accuracies across all classifiers dropped by $\approx 10\%$ compared to the in-domain case. The ECNN had the highest test accuracy, though like the in-domain case, the accuracies across the different models were not substantially different. The SDCNN was highest across all other metrics. More noticeably, it had the highest $D_{KL}$ by a substantial margin, indicating improved confidence estimation via better separation in the confidence scores of correct and incorrect decisions. This is particularly visible in Fig. \ref{fig:dd_distributions}, where the confidence score distribution of incorrect predictions were centered lower on the axis. All other models exhibited a higher concentration of high confidence predictions for both correct and incorrect ones.

\begin{figure*}[!t]
\begin{center}
    \includegraphics[width=1.5\columnwidth]{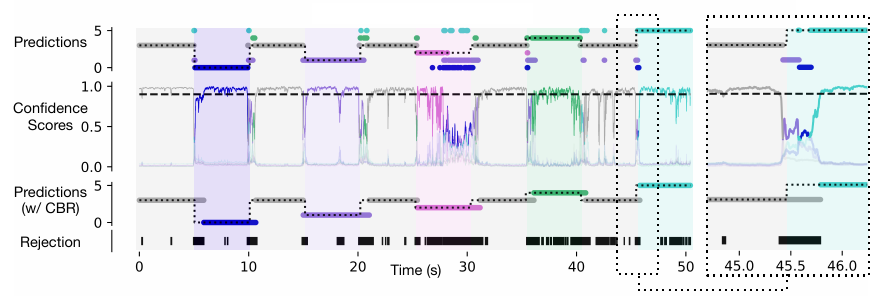}
    \includegraphics[width=1.5\columnwidth]{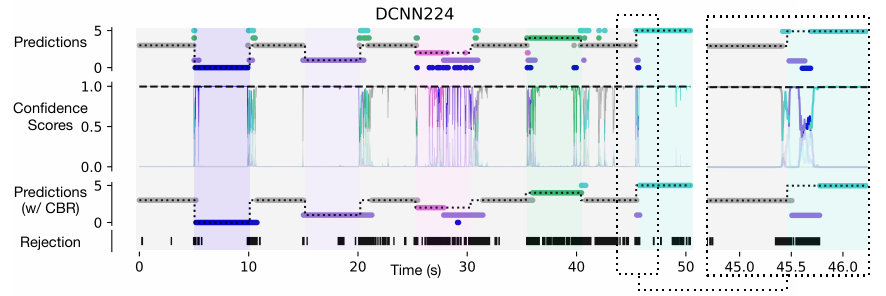}
    \includegraphics[width=1.5\columnwidth]{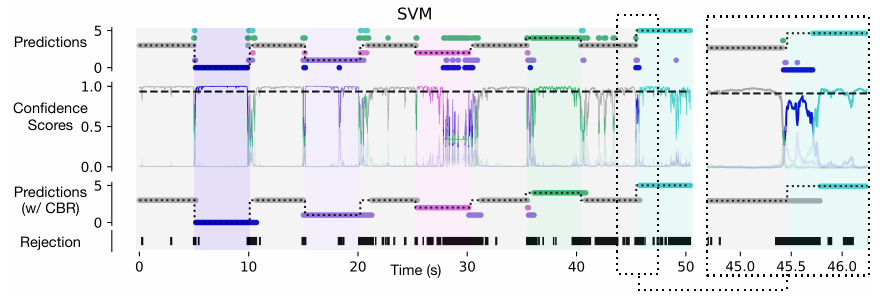}
    \includegraphics[width=1.5\columnwidth]{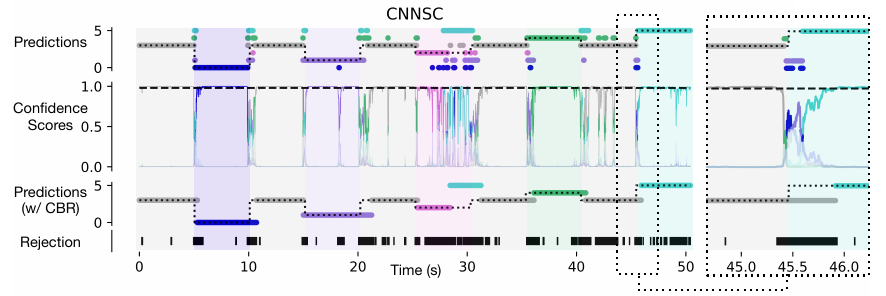}
    \includegraphics[width=1.5\columnwidth]{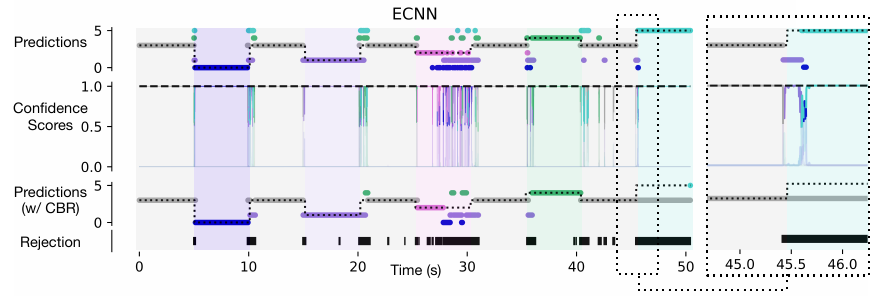}
\caption{Online predictions and confidence-based rejection (CBR). For each model, confidence scores were scaled to a common range of 0 to 1 for visualization purposes. The dotted lines and color overlays represent the target predictions. The confidence plots display the assigned score for every potential gesture prediction, with the top score depicted in solid color and the others in transparency.}
\label{fig:seq}
\end{center}
\end{figure*}

\subsection{Out-of-domain unforeseen class rejection}
The out-of-domain test evaluated the model behavior with unforeseen classes passed in the testing phase. With predictions being inherently wrong, only the accuracy from the in-domain reference testing is reported in Table \ref{tab:results}. From the CV folds, the out-of-domain test predictions are aggregated with the in-domain reference predictions, and relabeled as true or false based on whether they are in or out of the training domain. The confidence score distributions (Fig. \ref{fig:ood_distributions}) and the $D_{KL}$ (Table \ref{tab:results}) show how much the out-of-domain confidence distribution differs from the in-domain one.

Similar to the domain-divergent case, the SDCNN was highest across all models for most metrics, notably for our metric of interest for confidence estimation, the $D_{KL}$.

\subsection{Online decision rejection}
To provide a more detailed understanding of the models' online behavior, we selected one of the gesture sequence from the domain-divergent test to illustrate the corresponding predictions and confidence scores obtained in Fig. \ref{fig:seq}. We assessed their proficiency in evaluating and abstaining from uncertain predictions through a comparative analysis of decisions before and after rejection. For the comparison, the threshold was established so as to attain a consistent rejection rate across each model. Going up from 0 with 1\% increments, we stopped when one of the models first reaches a false positive rate (FPR) of 0. This process led to a rejection rate of 28\%, at which point the SDCNN had reached the 0 FPR. The SVM is the only other model able to reach a FPR of 0 on this sequence, but at a 48.98\% rejection rate. The other models were unable to do so at any rejection rate.

\begin{table*}[t]
  \caption{Experimental results}
  \label{tab:results}
  \centering
  \begin{tabular}{|lr||c|c|c|c|c|c|}
  \hline
    Experiment & Model     & Acc$_{test}$ & Acc$_{ref}$ & $D_{KL}$ & AUROC & AUPRC & AUARC\\  \hline \hline

    \textbf{In-domain}
    & SDCNN & 0.9385            & - & \textbf{1.7043}   & \textbf{0.9245}   & 0.9890            & \textbf{0.9901} \\
    & DCNN  & 0.9189            & - & 1.3155            & 0.8944            & 0.9847            & 0.9806 \\
    & SVM   & \textbf{0.9411}   & - & 1.6840            & 0.9220            & 0.9892            & 0.9897 \\
    & CNNSC & 0.9289            & - & 1.4890            & 0.9060            & 0.9871            & 0.9847 \\
    & ECNN  & 0.9373            & - & 1.3437            & 0.9156            & \textbf{0.9894}   & 0.9891 \\  \hline

    \textbf{Domain-divergent} 
    & SDCNN & 0.8174            & 0.9385            & \textbf{0.9229}   & \textbf{0.8598}   & \textbf{0.9581}   & \textbf{0.9464} \\
    & DCNN  & 0.7948            & 0.9189            & 0.6157            & 0.8029            & 0.9322            & 0.9108 \\
    & SVM   & 0.8053            & \textbf{0.9411}   & 0.5513            & 0.7806            & 0.9082            & 0.8972 \\
    & CNNSC & 0.8121            & 0.9289            & 0.6070            & 0.8257            & 0.9421            & 0.9232 \\
    & ECNN  & 0.\textbf{8289}   & 0.9373            & 0.5611            & 0.8393            & 0.9573            & 0.9401 \\  \hline

    \textbf{Out-of-domain} 
    & SDCNN & - & 0.9404            & \textbf{0.8338}   & \textbf{0.8138}   & \textbf{0.8001}   & \textbf{0.7154} \\
    & DCNN  & - & 0.9147            & 0.3445            & 0.7237            & 0.7083            & 0.6455 \\
    & SVM   & - & 0.9349            & 0.6015            & 0.7446            & 0.7149            & 0.6474 \\
    & CNNSC & - & 0.9376            & 0.5171            & 0.7703            & 0.7465            & 0.6792 \\
    & ECNN  & - & \textbf{0.9436}   & 0.4462            & 0.7550            & 0.7711            & 0.6666 \\  \hline

  \end{tabular}
\end{table*}

\begin{figure*}[t]
\begin{center}
    \includegraphics[width=1.7\columnwidth]{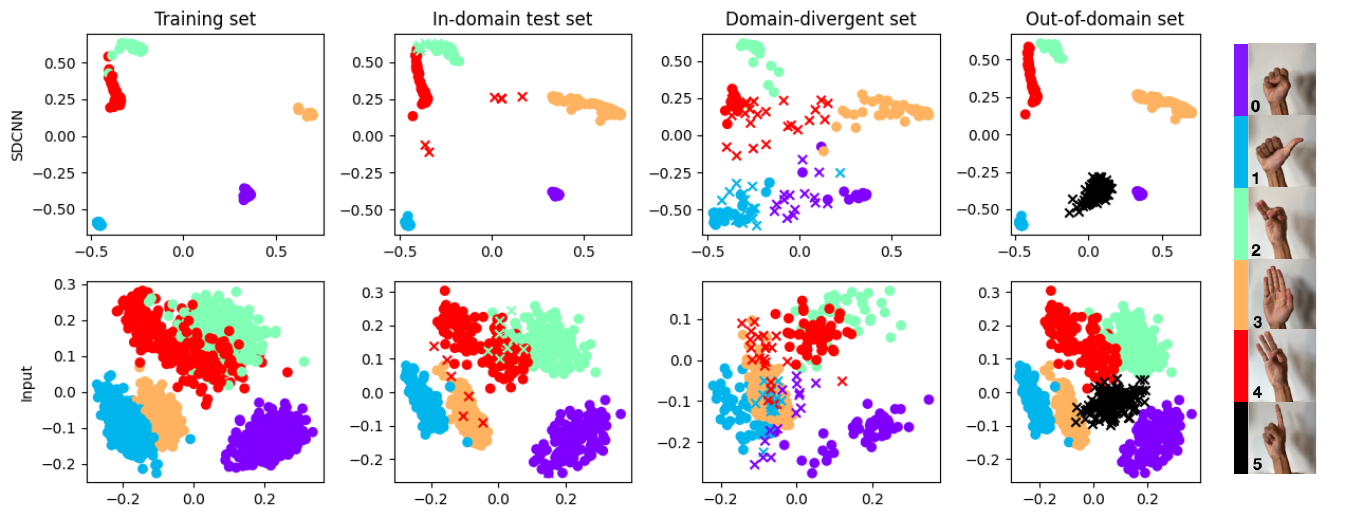}
\caption{Feature space visualization with PCA. Top row displays the SDCNN's 224-dimensional space projected onto the first 2 principal components. Bottom row displays the same for the input feature space. Correct predictions are labeled with an 'o' marker, incorrect ones with an 'x'.}
\label{fig:proj}
\end{center}
\end{figure*}

\subsection{Feature space visualization}
A deeper look into the model's embedding space further demonstrates its semantic reasoning and decision making. Shown in Fig. \ref{fig:proj} is one example of the training and test sets in the SDCNN and input feature spaces. To keep the comparison consistent across sets, the out-of-domain class was also left out from the training in-domain and domain-divergent sets. For visualization purpose, the data were linearly projected to 2 dimensions with principal component analysis (PCA), down from 224 and 64 (flattened 4x16 HD-EMG frames) for the SDCNN and input space, respectively. In each case, the PCA was fit exclusively with the training set. To improve visibility, the 45465 data points in the domain-divergent sequence set were downsampled to 456 (keeping 1 of every 100 predictions).
While there was already an apparent structure to the input space, noticeable re-arrangements can be seen in the model's embedding space, such as a better separation of classes 1 (thumb up) and 3 (rest). On the other hand, classes 2 (chuck grip) and 4 (pinch grip) remain closely located, reflecting their semantic similarity. Errors within the in-domain and domain-divergent sets occurred mostly between clusters in the continuous distance-based feature space, as expected. Confidence-based decision rejection can then discard these predictions on the basis of their increased distance to the nearest class centroid. Despite being unseen during training, the out-of-distribution class maintains its distinct cluster structure. This supports our hypothesis that meta-learning encourages learning generalizable patterns, i.e., based on data similarity rather than class labels.

\section{Discussion \& conclusion}
In this work, we explored the use of deep metric-based meta-learning as an alternative to the typical classification framework used in EMG PR applications. We developed a framework based on a Siamese network architecture trained using a contrastive triplet loss function, and combined it with centroid classification and decision rejection. We conducted a series of experiments to assess the approach's performance in comparison to a CNN trained with standard categorical cross-entropy, a non-DL SVM model, and two other solutions proposed in the field (CNNSC and ECNN). Our experiments show that despite fairly similar baseline accuracies between most models and test cases, the SDCNN provided better performance across the other metrics in almost all cases. The most notable difference was its consistently higher $D_{KL}$, which consequently yields better decision-rejection performance (AUARC). In other words, it was able to better identify domain-divergent and out of domain samples. This is further exemplified with sequential predictions and confidence estimation scores. The obtained results show that models trained with the metric learning framework yield better generalization, and supports the findings in the literature that conventional cross-entropy framework leads to poor generalization \cite{pmlr-v70-guo17a}. Our work also corroborates and brings EMG PR research up to speed with recent trends in the broader field of deep learning focusing on training backbone networks more specifically to produce useful representations for better performance in downstream tasks, in terms of generalization, transferability, or explainability for instance \cite{wang2023, pmlr-v119-chen20j}.

In addition, the output of the Siamese model holds human-level interpretable meaning as it gives proximity information between gesture classes, which other models do not inherently encode. This enables the identification of potentially confounding gestures, either with quantified contextual distances in the model's output, or by visualizing the feature space as in \ref{fig:proj}. With an intuitive EMG PR model that is more robust and transparent, our work sets the stage for a better integrated holistic myoelectric control solution. As suggested in \cite{jiang2023}, other key pieces for myoelectric neural interfaces are additional sensor modalities and sensory feedback. The latter will be particularly important to close the loop, in terms of human-in-the-loop bionic control systems, and achieve prosthetic limb embodiment \cite{farina2023}. In the same vein, our model builds towards more seamless bidirectional information transfer in the human-machine interface. The transparent and interpretable learned feature space can provide useful information to users and clinicians on neuromotor capabilities during the training process. Gesture cluster visualization could help improve contraction pattern consistency or indicate confounding gesture classes to omit as a control mode \cite{nawfel2022}.

At this time, our research has focused on \emph{offline} and post-hoc \emph{online} validation of our model, showing its performance in an open-loop system. The next steps are to integrate and fully leverage user performance when such a system is deployed for \emph{real-time} operation, where the user is closing the loop and can adapt dynamically to model predictions \cite{hahne2017}. In addition, this EMG metric-based meta-learning framework is amenable to further optimization through architectural or hyperparameter-oriented explorations.

In conclusion, this research advocates for a paradigm shift in deep learning for EMG PR, from classification frameworks to deep metric-based meta-learning. Proposing the utilization of a Siamese Deep Convolutional Neural Network (SDCNN) and contrastive triplet loss, the study shows its effectiveness for improved generalization and enhanced usability of EMG PR models in unconstrained domains. By focusing on learning meaningful and interpretable EMG feature representations, coupled with a confidence estimator based on class proximity, the research enables better rejection of incorrect decisions, which is especially effective when divering from the training data domain. This advancement paves the way for more reliable and practical EMG-based applications in real-world settings.

\section*{Acknowledgments}
We would like to thank the Microsystems Strategic Alliance of Québec (ReSMiQ), the Quebec Research Funds in Science and Technologies (FRQNT) and the Natural Sciences and Engineering Research Council of Canada (NSERC) for their support. The authors would also like to thank the Canada Research Chair in Smart Biomedical Microsystems and Bio6, our industrial partner.


\bibliographystyle{IEEEtran}
\bibliography{bibliography}

\begin{thebibliography}{10}
\providecommand{\url}[1]{#1}
\csname url@samestyle\endcsname
\providecommand{\newblock}{\relax}
\providecommand{\bibinfo}[2]{#2}
\providecommand{\BIBentrySTDinterwordspacing}{\spaceskip=0pt\relax}
\providecommand{\BIBentryALTinterwordstretchfactor}{4}
\providecommand{\BIBentryALTinterwordspacing}{\spaceskip=\fontdimen2\font plus
\BIBentryALTinterwordstretchfactor\fontdimen3\font minus \fontdimen4\font\relax}
\providecommand{\BIBforeignlanguage}[2]{{%
\expandafter\ifx\csname l@#1\endcsname\relax
\typeout{** WARNING: IEEEtran.bst: No hyphenation pattern has been}%
\typeout{** loaded for the language `#1'. Using the pattern for}%
\typeout{** the default language instead.}%
\else
\language=\csname l@#1\endcsname
\fi
#2}}
\providecommand{\BIBdecl}{\relax}
\BIBdecl

\bibitem{gopal2022}
P.~Gopal, A.~Gesta, and A.~Mohebbi, ``A systematic study on electromyography-based hand gesture recognition for assistive robots using deep learning and machine learning models,'' \emph{Sensors}, vol.~22, no.~10, p. 3650, 2022.

\bibitem{parajuli_real-time_2019}
N.~Parajuli, N.~Sreenivasan, P.~Bifulco, M.~Cesarelli, S.~Savino, V.~Niola, D.~Esposito, T.~J. Hamilton, G.~R. Naik, U.~Gunawardana, and G.~D. Gargiulo, ``\BIBforeignlanguage{en}{Real-{Time} {EMG} {Based} {Pattern} {Recognition} {Control} for {Hand} {Prostheses}: {A} {Review} on {Existing} {Methods}, {Challenges} and {Future} {Implementation}},'' \emph{\BIBforeignlanguage{en}{Sensors}}, vol.~19, no.~20, p. 4596, 2019.

\bibitem{jiang2023}
\BIBentryALTinterwordspacing
N.~Jiang, C.~Chen, J.~He, J.~Meng, L.~Pan, S.~Su, and X.~Zhu, ``{Bio-robotics research for non-invasive myoelectric neural interfaces for upper-limb prosthetic control: a 10-year perspective review},'' \emph{National Science Review}, vol.~10, no.~5, p. nwad048, 02 2023. [Online]. Available: \url{https://doi.org/10.1093/nsr/nwad048}
\BIBentrySTDinterwordspacing

\bibitem{campbell2020}
E.~Campbell, A.~Phinyomark, and E.~Scheme, ``Current trends and confounding factors in myoelectric control: Limb position and contraction intensity,'' \emph{Sensors}, vol.~20, no.~6, p. 1613, 2020.

\bibitem{robertson2019}
J.~W. Robertson, K.~B. Englehart, and E.~J. Scheme, ``Effects of confidence-based rejection on usability and error in pattern recognition-based myoelectric control,'' \emph{IEEE Journal of Biomedical and Health Informatics}, vol.~23, no.~5, pp. 2002--2008, 2019.

\bibitem{scheme2015}
E.~Scheme and K.~Englehart, ``A comparison of classification based confidence metrics for use in the design of myoelectric control systems,'' in \emph{2015 37th Annual International Conference of the IEEE Engineering in Medicine and Biology Society (EMBC)}, 2015, pp. 7278--7283.

\bibitem{shri2023}
S.~T.~P. Raghu, D.~MacIsaac, and E.~Scheme, ``Decision-change informed rejection improves robustness in pattern recognition-based myoelectric control,'' \emph{IEEE Journal of Biomedical and Health Informatics}, vol.~27, no.~12, pp. 6051--6061, 2023.

\bibitem{tjoa2021}
E.~Tjoa and C.~Guan, ``A survey on explainable artificial intelligence (xai): Toward medical xai,'' \emph{IEEE Transactions on Neural Networks and Learning Systems}, vol.~32, no.~11, pp. 4793--4813, 2021.

\bibitem{pmlr-v70-guo17a}
C.~Guo, G.~Pleiss, Y.~Sun, and K.~Q. Weinberger, ``On calibration of modern neural networks,'' in \emph{Proceedings of the 34th International Conference on Machine Learning}, ser. Proceedings of Machine Learning Research, D.~Precup and Y.~W. Teh, Eds., vol.~70.\hskip 1em plus 0.5em minus 0.4em\relax PMLR, 06--11 Aug 2017, pp. 1321--1330.

\bibitem{NEURIPS2019_36ad8b5f}
S.~Thulasidasan, G.~Chennupati, J.~A. Bilmes, T.~Bhattacharya, and S.~Michalak, ``On mixup training: Improved calibration and predictive uncertainty for deep neural networks,'' in \emph{Advances in Neural Information Processing Systems}, H.~Wallach, H.~Larochelle, A.~Beygelzimer, F.~d\textquotesingle Alch\'{e}-Buc, E.~Fox, and R.~Garnett, Eds., vol.~32.\hskip 1em plus 0.5em minus 0.4em\relax Curran Associates, Inc., 2019.

\bibitem{jaramillo2020}
A.~Jaramillo-Y{\'a}nez, M.~E. Benalc{\'a}zar, and E.~Mena-Maldonado, ``Real-time hand gesture recognition using surface electromyography and machine learning: a systematic literature review,'' \emph{Sensors}, vol.~20, no.~9, p. 2467, 2020.

\bibitem{tam2021}
S.~Tam, M.~Boukadoum, A.~Campeau-Lecours, and B.~Gosselin, ``Intuitive real-time control strategy for high-density myoelectric hand prosthesis using deep and transfer learning,'' \emph{Scientific Reports}, vol.~11, no.~1, pp. 1--14, 2021.

\bibitem{cote2019}
U.~C{\^o}t{\'e}-Allard, C.~L. Fall, A.~Drouin, A.~Campeau-Lecours, C.~Gosselin, K.~Glette, F.~Laviolette, and B.~Gosselin, ``Deep learning for electromyographic hand gesture signal classification using transfer learning,'' \emph{IEEE transactions on neural systems and rehabilitation engineering}, vol.~27, no.~4, pp. 760--771, 2019.

\bibitem{cote2021}
U.~C{\^o}t{\'e}-Allard, G.~Gagnon-Turcotte, A.~Phinyomark, K.~Glette, E.~Scheme, F.~Laviolette, and B.~Gosselin, ``A transferable adaptive domain adversarial neural network for virtual reality augmented emg-based gesture recognition,'' \emph{IEEE Transactions on Neural Systems and Rehabilitation Engineering}, vol.~29, pp. 546--555, 2021.

\bibitem{du2017}
Y.~Du, W.~Jin, W.~Wei, Y.~Hu, and W.~Geng, ``{Surface EMG-based inter-session gesture recognition enhanced by deep domain adaptation},'' \emph{Sensors}, vol.~17, no.~3, p. 458, 2017.

\bibitem{tsinganos2020}
P.~Tsinganos, B.~Cornelis, J.~Cornelis, B.~Jansen, and A.~Skodras, ``Data augmentation of surface electromyography for hand gesture recognition,'' \emph{Sensors}, vol.~20, no.~17, p. 4892, 2020.

\bibitem{chamberland2023}
F.~Chamberland, E.~Buteau, S.~Tam, E.~Campbell, A.~Mortazavi, E.~Scheme, P.~Fortier, M.~Boukadoum, A.~Campeau-Lecours, and B.~Gosselin, ``Novel wearable hd-emg sensor with shift-robust gesture recognition using deep learning,'' \emph{IEEE Transactions on Biomedical Circuits and Systems}, vol.~17, no.~5, pp. 968--984, 2023.

\bibitem{bao2022}
T.~Bao, S.~A.~R. Zaidi, S.~Q. Xie, P.~Yang, and Z.-Q. Zhang, ``Cnn confidence estimation for rejection-based hand gesture classification in myoelectric control,'' \emph{IEEE Transactions on Human-Machine Systems}, vol.~52, no.~1, pp. 99--109, 2022.

\bibitem{lin2022}
Y.~Lin, R.~Palaniappan, P.~De~Wilde, and L.~Li, ``Reliability analysis for finger movement recognition with raw electromyographic signal by evidential convolutional networks,'' \emph{IEEE Transactions on Neural Systems and Rehabilitation Engineering}, vol.~30, pp. 96--107, 2022.

\bibitem{wu2021}
L.~Wu, X.~Zhang, X.~Zhang, X.~Chen, and X.~Chen, ``Metric learning for novel motion rejection in high-density myoelectric pattern recognition,'' \emph{Knowl. Based Syst.}, vol. 227, p. 107165, 2021.

\bibitem{huisman2021}
\BIBentryALTinterwordspacing
M.~Huisman, J.~N. van Rijn, and A.~Plaat, ``\BIBforeignlanguage{en}{A survey of deep meta-learning},'' \emph{\BIBforeignlanguage{en}{Artificial Intelligence Review}}, vol.~54, no.~6, pp. 4483--4541, Aug. 2021. [Online]. Available: \url{https://doi.org/10.1007/s10462-021-10004-4}
\BIBentrySTDinterwordspacing

\bibitem{prorokovi2020}
\BIBentryALTinterwordspacing
K.~Prorokovi{\'c}, M.~Wand, and J.~Schmidhuber, ``Meta-learning for recalibration of {EMG}-based upper limb prostheses,'' in \emph{4th Lifelong Machine Learning Workshop at ICML 2020}, 2020. [Online]. Available: \url{https://openreview.net/forum?id=wRI-iDtHLoM}
\BIBentrySTDinterwordspacing

\bibitem{rahimian2021}
E.~Rahimian, S.~Zabihi, A.~Asif, D.~Farina, S.~F. Atashzar, and A.~Mohammadi, ``Fs-hgr: Few-shot learning for hand gesture recognition via electromyography,'' \emph{IEEE Transactions on Neural Systems and Rehabilitation Engineering}, vol.~29, pp. 1004--1015, 2021.

\bibitem{fan2022}
\BIBentryALTinterwordspacing
X.~Fan, L.~Zou, Z.~Liu, Y.~He, L.~Zou, and R.~Chi, ``Csac-net: Fast adaptive semg recognition through attention convolution network and model-agnostic meta-learning,'' \emph{Sensors}, vol.~22, no.~10, 2022. [Online]. Available: \url{https://www.mdpi.com/1424-8220/22/10/3661}
\BIBentrySTDinterwordspacing

\bibitem{tam2019}
S.~Tam, M.~Boukadoum, A.~Campeau-Lecours, and B.~Gosselin, ``A fully embedded adaptive real-time hand gesture classifier leveraging hd-semg and deep learning,'' \emph{IEEE transactions on biomedical circuits and systems}, vol.~14, no.~2, pp. 232--243, 2019.

\bibitem{krizhevsky2017}
A.~Krizhevsky, I.~Sutskever, and G.~E. Hinton, ``Imagenet classification with deep convolutional neural networks,'' \emph{Communications of the ACM}, vol.~60, no.~6, pp. 84--90, 2017.

\bibitem{schroff2015}
\BIBentryALTinterwordspacing
F.~Schroff, D.~Kalenichenko, and J.~Philbin, ``Facenet: {A} unified embedding for face recognition and clustering,'' \emph{CoRR}, vol. abs/1503.03832, 2015. [Online]. Available: \url{http://arxiv.org/abs/1503.03832}
\BIBentrySTDinterwordspacing

\bibitem{wang2022}
R.-Q. Wang, X.-Y. Zhang, and C.-L. Liu, ``Meta-prototypical learning for domain-agnostic few-shot recognition,'' \emph{IEEE Transactions on Neural Networks and Learning Systems}, vol.~33, no.~11, pp. 6990--6996, 2022.

\bibitem{nguyen2015}
A.~Nguyen, J.~Yosinski, and J.~Clune, ``Deep neural networks are easily fooled: High confidence predictions for unrecognizable images,'' in \emph{Proceedings of the IEEE Conference on Computer Vision and Pattern Recognition (CVPR)}, June 2015.

\bibitem{phinyomark2013}
A.~Phinyomark, F.~Quaine, S.~Charbonnier, C.~Serviere, F.~Tarpin-Bernard, and Y.~Laurillau, ``Emg feature evaluation for improving myoelectric pattern recognition robustness,'' \emph{Expert Systems with Applications}, vol.~40, pp. 4832--4840, 09 2013.

\bibitem{shri2022}
\BIBentryALTinterwordspacing
S.~{Tallam Puranam Raghu}, D.~MacIsaac, and E.~Scheme, ``Analyzing the impact of class transitions on the design of pattern recognition-based myoelectric control schemes,'' \emph{Biomedical Signal Processing and Control}, vol.~71, p. 103134, 2022. [Online]. Available: \url{https://www.sciencedirect.com/science/article/pii/S174680942100731X}
\BIBentrySTDinterwordspacing

\bibitem{li2021}
W.~Li, P.~Shi, and H.~Yu, ``Gesture recognition using surface electromyography and deep learning for prostheses hand: State-of-the-art, challenges, and future,'' \emph{Frontiers in neuroscience}, vol.~15, p. 621885, 2021.

\bibitem{wang2023}
W.~Wang, C.~Han, T.~Zhou, and D.~Liu, ``Visual recognition with deep nearest centroids,'' 2023.

\bibitem{pmlr-v119-chen20j}
\BIBentryALTinterwordspacing
T.~Chen, S.~Kornblith, M.~Norouzi, and G.~Hinton, ``A simple framework for contrastive learning of visual representations,'' in \emph{Proceedings of the 37th International Conference on Machine Learning}, ser. Proceedings of Machine Learning Research, H.~D. III and A.~Singh, Eds., vol. 119.\hskip 1em plus 0.5em minus 0.4em\relax PMLR, 13--18 Jul 2020, pp. 1597--1607. [Online]. Available: \url{https://proceedings.mlr.press/v119/chen20j.html}
\BIBentrySTDinterwordspacing

\bibitem{farina2023}
D.~Farina, I.~Vujaklija, R.~Br{\aa}nemark, A.~M. Bull, H.~Dietl, B.~Graimann, L.~J. Hargrove, K.-P. Hoffmann, H.~Huang, T.~Ingvarsson \emph{et~al.}, ``Toward higher-performance bionic limbs for wider clinical use,'' \emph{Nature biomedical engineering}, vol.~7, no.~4, pp. 473--485, 2023.

\bibitem{nawfel2022}
J.~L. Nawfel, K.~B. Englehart, and E.~J. Scheme, ``The influence of training with visual biofeedback on the predictability of myoelectric control usability,'' \emph{IEEE Transactions on Neural Systems and Rehabilitation Engineering}, vol.~30, pp. 878--892, 2022.

\bibitem{hahne2017}
J.~M. Hahne, M.~Markovic, and D.~Farina, ``User adaptation in myoelectric man-machine interfaces,'' \emph{Scientific Reports}, vol.~7, p. 4437, 12 2017.

\end{thebibliography}

% \newpage

 

% \end{IEEEbiography}



\vfill

\end{document}